\documentclass[prb,twocolumn,showpacs,preprintnumbers,amsmath,amssymb,showkeys]{revtex4}

\usepackage{graphicx}
\usepackage{bm}

\begin{document}


\title{Microwave response of thin niobium films under perpendicular static magnetic fields}

\author{D. Janju\v{s}evi\'c}
\author{M. S. Grbi\'c}
\author{M. Po\v{z}ek}
\email{mpozek@phy.hr} 
\author{A. Dul\v{c}i\'{c}}
\author{D. Paar}
\affiliation{Department of Physics, Faculty of Science, University of Zagreb, P. O. Box 331,
HR-10002 Zagreb, Croatia}
\author{B. Nebendahl}\thanks{present address: Agilent Technologies R\&D and Marketing GmbH \&  
Co. KG, Herrenberger Strasse 130, D-71034 B\"oblingen, Germany}
\affiliation
{2. Physikalisches Institut, Universit\"at Stuttgart,
D-70550 Stuttgart, Germany}
\author{T. Wagner}
\affiliation
{%
Max-Planck-Institute for Metals Research, 
Heisenbergstrasse 3, D-70156 Stuttgart, Germany
}%


\begin{abstract}
The microwave response of high quality niobium films in a perpendicular static magnetic field has been investigated. 
The complex frequency shift was measured up to the
upper critical fields. The data have been analyzed by the effective conductivity model for the 
type-II superconductors in the mixed state.
This model is found to yield consistent results for the coherence lengths in high-$\kappa$ 
superconducting samples, and can be used with HTSC even at temperatures much below $T_c$. It is shown that for 
samples with high values of depinning frequency, one should measure both components of the complex frequency shift
in order to determine the flow resistivity.

The thick Nb film (160\:nm) has low resistivity at 10 K, comparable to the best single crystals, and low $\kappa$
value. In contrast, the thinnest (10\:nm) film has $\kappa \approx 9.5$ and exhibits a high depinning frequency
($\approx 20\:GHz$).
The upper critical field determined from microwave measurements is related to the radius of nonoverlaping vortices, 
and appears to be larger than the one determined by the transition to the normal state.
\end{abstract}

\maketitle

\section{Introduction} \label{sec:level1}

Microwave investigations of classical superconductors half a century ago have proven to be a very useful tool in the determination
of intrinsic parameters in the Meissner state\cite{Glover:57,Biondi:59,Klein:94} as well as the flow resistivity in the mixed 
state.\cite{Cardona:64,Rosenblum:64,Gittleman:66}
These investigations were also of large technical interest for the construction of microwave transmission lines 
and resonant structures.
In the last twenty years high temperature superconductors (HTSC) have been extensively investigated by microwave techniques. 
Among the greatest successes, one should 
mention the determination of the temperature dependence of the penetration depth which evidenced the existence of nodes in the 
superconducting gap, leading to the d-wave explanation of HTSC.\cite{Hardy:93}

Flux flow resistivity can be determined from dc measurements when the current density $J$ 
exceeds the critical current density $J_c$ for vortex depinning.\cite{Strnad:64} 
However, in many cases of low and high temperature superconductors, the dynamics of
vortex motion is more complicated and cannot be interpreted just in terms of pinned and depinned regimes.
Thermally activated flux hopping may yield ohmic behavior even for current densities below $J_c$.\cite{Palstra:90}
However, the resistivity observed in this regime should not be identified with that of flux flow. 
By increasing the current density there occurs a nonlinear
transition to the flux flow regime. At much higher current densities one may encounter
nonlinear effects in the flux flow leading ultimately to the point of instability where a dramatic
increase of the measured voltage occurs.\cite{Klein:85,Doettinger:94,Xiao:96,Ruck:97,Kunchur:02,Babic:04}
The nonlinear behavior is temperature and magnetic field dependent.
Hence, the linear flux flow regime may occur inbetween the two nonlinear regimes, and may not always be well resolved.

In contrast to dc resistivity, microwave measurements can be carried out with current densities much smaller than $J_c$, and still yield
flux flow resistivity. Namely, the vortices are not driven over the pinning potential barrier, but just 
oscillate within the potential well.

Depinning frequency $\omega_0$ separates two regimes of vortex oscillation. If the driving frequency $\omega$
is much larger than $\omega_0$, the viscous drag force dominates over the restoring pinning force in the response of 
vortices to the microwave Lorentz force.
This is often the case with classical superconductors which have $\omega_0$ in the MHz range.
In that case, the flux flow resistivity can be extracted from the microwave absorption curves only.
When $\omega_0 \approx \omega$, or higher, the restoring pinning force is comparable to the viscous drag force,
and the response of the vortices is more complex. This is typical of HTSC where the depinning frequency
is found to be of the order of 10 GHz,\cite{Golosovsky:94} but may be found also in low temperature superconductors
such as in very thin Nb film analyzed in this paper below. 
In such cases, one needs both, microwave absorption and dispersion to determine the depinning frequency and flux flow parameters.
%
%
The analysis usually employs the models 
of effective conductivity in the mixed state\cite{Coffey:91,Brandt:91,Dulcic:93} based on the Bardeen-Stephen model.\cite{Bardeen:65} 

The real values of $B_{c2}$ in HTSC remain experimentally unreachable except for the narrow temperature range below $T_c$, and 
the $B_{c2}$ values extracted from the effective conductivity data could not be experimentally verified. 
In order to probe the effective conductivity model, we have performed a series of measurements on thin films and a single
crystal of niobium, a classical type-II superconductor. The comparison of thick niobium samples with HTSC samples is not ideal
since the upper critical fields and depinning frequencies in niobium are much lower and the vortex distance doesn't always allow
the use of effective conductivity models. However, as the film thickness is reduced, one expects considerable enhancement
of $B_{c2}$, due to the reduced coherence length, and much stronger pinning, due to the surface effects.  

We have measured high quality niobium thin 
films using a cavity perturbation method. The temperature and field dependence of the complex frequency shift have been
measured up to the upper critical fields in order to test the validity of the effective conductivity models. The results of
the present analysis should be taken into account when the mixed state of HTSC is investigated by the microwave methods.

\section{Samples} \label{sec:level2}

The Nb films were deposited via molecular beam epitaxy (MBE) in a commercial system from DCA instruments (Finland). 
The base pressure of the system is 10$^{-9}$ Pa. The (0001) surfaces of the sapphire substrates ($\alpha$-Al$_2$O$_3$) were prepared by 
sputter cleaning with Ar ions (1 keV) and subsequent annealing at 1000$^{\circ}$C in ultra-high vacuum (UHV). 
This annealing procedure is necessary to remove the embedded Ar gas atoms and to recover the sapphire surface crystallography and morphology. 
Details of the surface preparation procedure can be found elsewhere.\cite{Bernath:98,Gao:02,Wagner:98} 
The substrate treatment resulted in unreconstructed surfaces, as revealed by reflection high energy electron diffraction (RHEED). 
Niobium (4 N purity) was evaporated from an electron beam 
evaporator (substrate temperature 900$^{\circ}$C; film thicknesses 10\:nm, 40\:nm, 160\:nm) and the growth rate was monitored using a quartz oscillator. 
Typical growth rates were between 0.01 and 0.05 nm/s. As revealed by in-situ RHEED investigations, the Nb films grew epitaxially on the (0001) sapphire 
substrate.\cite{Wagner:98} A 2\:nm thick protective layer of SiO$_2$ has been deposited onto each film 
in order to prevent niobium oxidation (physical vapour deposition at room temperature). 
Transmission electron microscopy (TEM) investigations have shown that the film of nominal thickness 
40\:nm is actually 36\:nm thick, indicating that the other films are also slightly thinner than their nominal thickness. 
The samples were cut by a diamond saw into pieces 3$\times$0.5 mm$^2$ (40\:nm and 160\:nm films), 
and 2.3$\times$1 mm$^2$ (10\:nm film), suitable for cavity perturbation measurements. 
High purity single crystal of niobium was purchased from Metal Crystals \& Oxides Ltd., Cambridge. 
Its dimensions were 3$\times$2$\times$0.5\:mm$^3$.

\begin{figure}
\includegraphics[width=8cm]{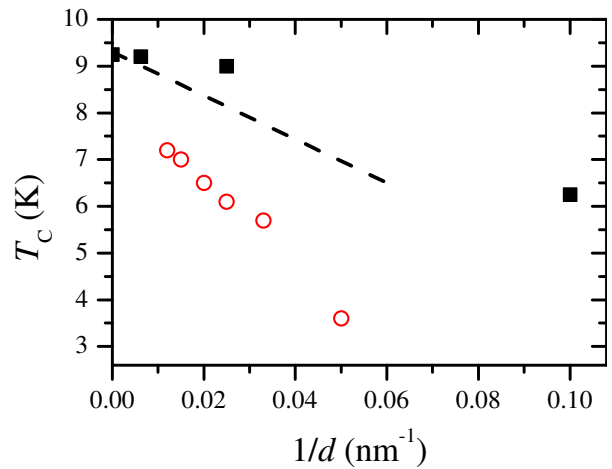}%
\caption{(Color online) Thickness dependence of the transition temperature of the measured niobium films (full squares). 
The data are compared with clean samples of Wolf et al.\cite{Wolf:76} and Kodama et al.\cite{Kodama:83} (dashed line),
and with dirty samples of Minhaj et al.\cite{Minhaj:94} (circles).}
\label{Fig1}
\end{figure}

Superconducting properties of thin niobium films strongly depend on their thickness, purity and preparation 
conditions.\cite{Wolf:76,Kodama:83,Minhaj:94,Park:85,Gubin:05,Hsu:92}
In order to compare our samples with those measured earlier,\cite{Wolf:76,Kodama:83,Minhaj:94}
we have determined T$_c$ of each film by the so called 90\% criterion, i.e. the transition temperature was defined as the 
temperature where microwave absorption $\Delta (1/2Q)$ reaches 90\% of its normal state value. For the samples of the same composition and purity
one expects that T$_c$ is reduced aproximately with the inverse thickness $d^{-1}$ due to the proximity effects,\cite{Cooper:61}
weak localization and increased residual resistivity.\cite{Minhaj:94} We show in Fig.\ref{Fig1} the comparison of 
transition temperatures of our thin films with results of some other authors. The $T_c$ values of present films are similar to the films measured
by Gubin et al.,\cite{Gubin:05} and, if extrapolated to ultrathin 2 nm, with the film measured by Hsu and Kapitulnik.\cite{Hsu:92} 
Obviously, the films measured in this paper are of good quality. 

To extract various superconducting parameters one needs three quantities: $T_c$, $\rho_n$ and $S=-dB_{c2}/dT$ at $T_c$. 
The $B_{c2}$ values determined also by the 90\% criterion 
from our measurements
gave the $S$-values $0.47\:\rm{T/K}$, $0.11\:\rm{T/K}$ 
and $0.08\:\rm{T/K}$ for 10\:nm, 40\:nm and 160\:nm samples, respectively.

The resistivities of the films have been determined by standard four contact method. Their residual values at 
$T=10\: \rm{K}$ were $\rho_n(10\:\rm{nm})=15.2\: \rm{\mu \Omega cm}$, $\rho_n(40\:\rm{nm})=1.4\: \rm{\mu \Omega cm}$ and 
$\rho_n(160\:\rm{nm})=0.33\: \rm{\mu \Omega cm}$, respectively.
The residual resistivity of 160\:nm film was 
found to be an order of magnitude lower than in the film studied by Gubin et al.,\cite{Gubin:05} and
comparable to the best bulk single crystals,\cite{Blaschke:82,Klein:94} showing that the 
films grew with perfect epitaxy with virtually no defects. 
The residual resistivities of thinner samples are, therefore, increased only due to the surface scattering, but the values are still below the 
values reported by Gubin et al.\cite{Gubin:05} For the 10\:nm film we estimate $\kappa \approx 9.5$, for the 40\:nm film 
$\kappa \approx 1.5$, and for the 160\:nm film we can take the single crystal value of $\kappa \approx 0.9$. 


\section{Experimental Details} \label{sec:level3}

Microwave measurements were carried out in a high-Q elliptical cavity made of copper resonating in  $_e$TE$_{111}$ 
mode at 9.3 GHz or in $_e$TE$_{113}$ mode at 17.5 GHz. 
For both modes used, microwave magnetic field has node in the center of the cavity while microwave electric field is at maximum. 
The sample was mounted on a sapphire sample holder and placed in the 
center of the cavity where the microwave electric field $E_{mw}$ has its maximum in both modes. 
The sample was oriented with the longest side parallel to $\boldsymbol{E}_{mw}$.
Experimental setup included Oxford Systems superconducting magnet with $\boldsymbol{B}_{dc}\perp \boldsymbol{E}_{mw}$.
While $\boldsymbol{E}_{mw}$ is always in the film plane, $\boldsymbol{B}_{dc}$ can be in the film plane or perpendicular to it.
Directly measured quantities are the $Q$-factor and the resonant 
frequency $f$ of the cavity loaded with the sample. 
The $Q$-factor was measured by a modulation technique described 
elsewhere.\cite{Nebendahl:01} The empty cavity absorption 
$(1/2Q)$ was substracted from the measured data and the presented 
experimental curves are due to the samples themselves.
An automatic frequency control (AFC) system was used to set the
source frequency always in resonance with the cavity. Thus, the
frequency shift can be measured as the temperature of the sample or static magnetic field is
varied. 
The two measured quantities represent the complex frequency shift 
$\Delta \widetilde{\omega} / \omega = \Delta f/f  + i \Delta (1/2Q)$.

\section{Effective conductivity} \label{sec:level4}

For the extraction of complex conductivity data from the measured complex frequency shift
of a thin superconducting sample in the microwave electric field, one can utilize
the general solution for the complex frequency shift by Peligrad et al.\cite{Peligrad:01}. 
The shift from a perfect conductor state is given by
\begin{equation}
\displaystyle {\frac{\Delta \widetilde{\omega}_p}{\omega}}=
\frac{\Gamma}{N} \left[
1+\left(\frac{\widetilde{k}^2}{k_0^2} 
\frac{\tanh (i \widetilde{k} d/2) }{i \widetilde{k} d/2} -1 \right) N
\right]^{-1}
\mbox{\ ,} 
\label{eq:2}
\end{equation}
where $d$ is the thickness of the slab, $N$ is the depolarization factor and 
$\Gamma$ is the dimensionless filling factor of the sample in the cavity. 
The complex wave vector $\widetilde{k}$ is given by
\begin{equation}
\widetilde{k}=k_0 \sqrt{\widetilde{\mu}_r \left( \widetilde{\epsilon}_r -
i \frac{\widetilde{\sigma}}{\epsilon_0 \omega} \right) } \mbox{\ ,} 
\label{eq:3}
\end{equation}
where $k_0=\omega \sqrt{\mu_0 \epsilon_0}$ is the vacuum wave vector. It
describes generally any set of material parameters. For a nonmagnetic metal one can take 
$\widetilde{\mu_r}=\widetilde{\epsilon_r}=1$ and the main contribution to $\widetilde{k}$ comes from conductivity.
Using these equations one can by numerical inversion of experimentally obtained complex frequency shift 
determine the complex conductivity $\widetilde{\sigma}$ of the sample.

For a superconducting sample in the mixed state, in intermediate fields one can define
the effective complex conductivity\cite{Coffey:91,Brandt:91,Dulcic:93} which is a combination 
of normal conductivity in the vortex cores and the conductivity of the condensed electrons 
outside the cores. The effective conductivity in an oscillating electric field is given by:
\begin{figure*}
\includegraphics[width=18cm]{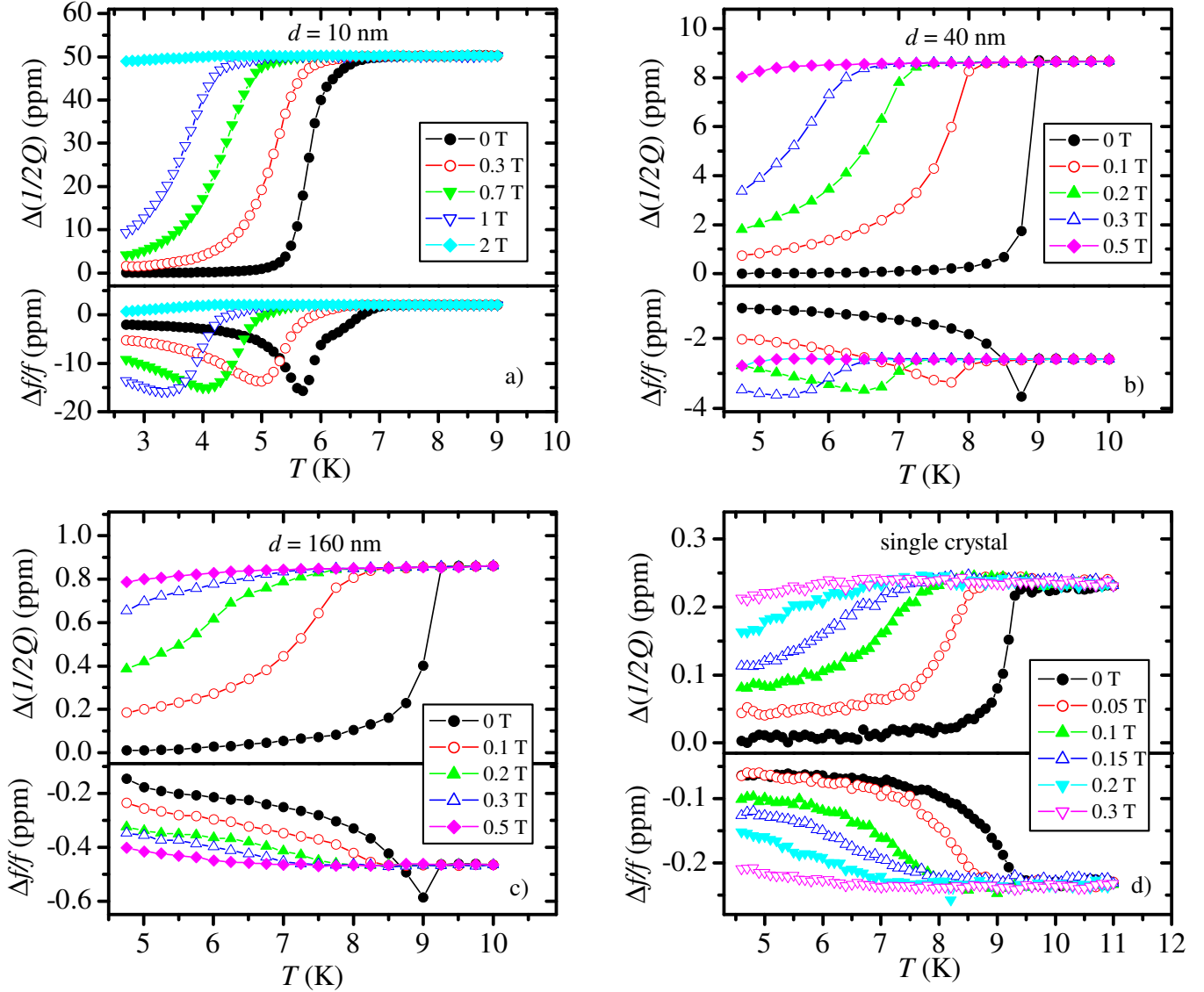}%
\caption{(Color online) Measured temperature dependence of complex frequency shift of four niobium samples: 
a) thin film $d=10\:\rm{nm}$; b) thin film $d=40\:\rm{nm}$; c) thin film $d=160\:\rm{nm}$; d) single crystal.
Driving frequency was 9.3\:GHz. Applied perpendicular static magnetic fields are indicated in the legends.}
\label{Fig2}
\end{figure*}
%

\begin{equation}
\frac{1}{\widetilde{\sigma}_{\mathrm{eff}}}= \frac{1-\frac{b}{1-i(\omega_0/\omega)}}
{(1-b)(\sigma_1-i \sigma_2)+b \sigma_n}+
\frac{1}{\sigma_n}\, \frac{b}{1-i(\omega_0/\omega)}
\label{eq:4}
\end{equation}
The first term is due to the microwave current outside the vortex cores, and the second is due to 
the normal current in the cores of the oscillating vortices. The meaning of the parameter 
$b$ in Eq.~(\ref{eq:4}) is the volume fraction of the sample taken by the normal vortex 
cores. This parameter determines the resistivity in the flux flow regime $\rho_f / \rho_n$.\cite{Tinkham:96} 
The depinning frequency $\omega_0$ may vary, depending on sample, field and temperature from 
strongly pinned case ($\omega_0  \gg \omega$) to the flux flow limit ($\omega_0 \ll \omega$). 
In Eq.~(\ref{eq:4}) the zero field conductivity is $\sigma_1 - i \sigma_2$, and $\sigma_n$ is 
the normal state conductivity. The model is limitted to high $\kappa$-values and to magnetic fields
much lower than upper critical fields, where vortices don't overlap significantly.

From the experimentally obtained field dependent complex conductivity extracted 
using Eqs.~(\ref{eq:2}) and (\ref{eq:3}) 
one can, by numerical inversion of Eq.~(\ref{eq:4}), determine the 
values of $b$ and $\omega_0 / \omega$.

\section{Results and Discussion} \label{sec:level6}

The measured complex frequency shift of the three films and the single crystal is presented in Fig.~\ref{Fig2}
for various values of DC magnetic field and for a driving frequency 9.3\:GHz. 
One may notice a huge difference in signal intensities between 
thin and thick samples due to very different filling factors $\Gamma$. 
\begin{figure}
\includegraphics[width=8cm]{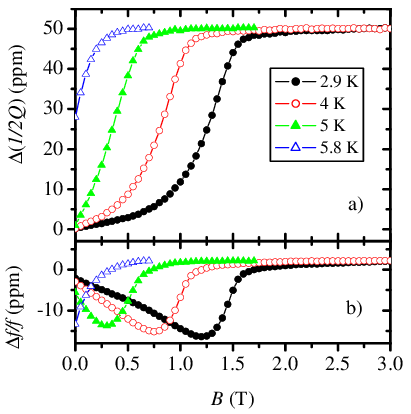}%
\caption{(Color online) Field dependence of the complex frequency shift for the 10\:nm film at various temperatures.}
\label{Fig3}
\end{figure}
Namely, a conducting sample profoundly 
changes the electric field at the centre of the cavity with respect to the empty cavity field. The sample acts as a partial
short for the electric field lines.\cite{Peligrad:98} The field outside of a conductor is stronger at 
rear and front sides, and this enhancement is larger for thinner films. As a consequence, the microwave
measurements in electric field are very sensitive for thin films, while for thicker samples
the signal/noise ratio is reduced. 

Looking at the zero field measurements in Fig.~\ref{Fig2}, one observes a smooth transition to the 
superconducting state for the thinnest film, and sharp transitions for thicker samples. In the 10\:nm sample
the fluctuation effects become considerable since the thickness becomes lower than the coherence length.

From the measured frequency shift and known normal state resistivity $\rho_n$ one can 
determine the 
complex conductivity in the whole temperature range and consequently the zero temperature penetration depth.
We have 
obtained 
$\lambda(T=0)=285\:\rm{nm}$ for the 10\:nm film, comparable to the result of Gubin et 
al.\cite{Gubin:05} for the film of the same thickness. 
%
%
The filling factor $\Gamma$ was determined at $T=10\: \rm{K}$ from the measured frequency shift and known 
normal state resistivity for each sample, and we keep it fixed in further analysis of a given sample.

\begin{figure}
\includegraphics[width=8cm]{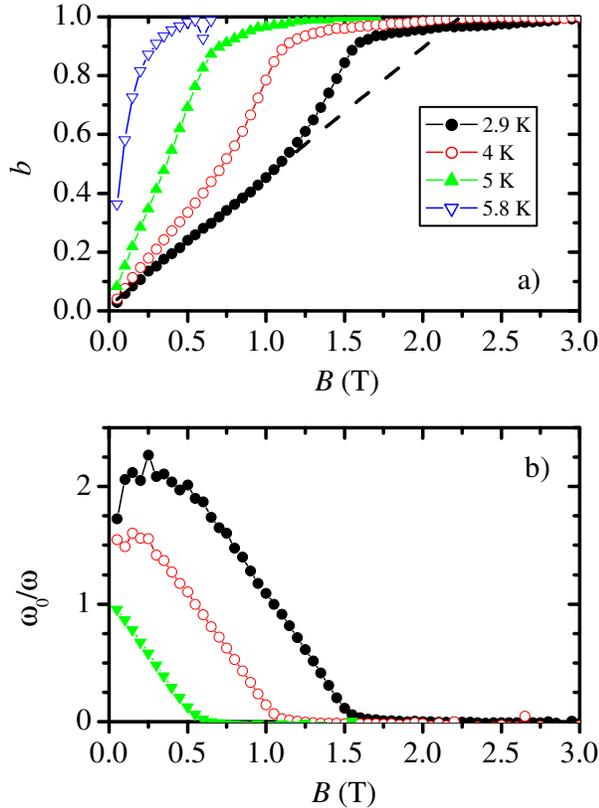}%
\caption{(Color online) Parameters $b$ and $\omega_0 / \omega$ of the effective conductivity model extracted from the data of Fig.~\ref{Fig3}.
The dashed line shows the extrapolation of the linear section at $T=2.9\:\rm{K}$.}
\label{Fig4}
\end{figure}
Knowing the measured temperature dependence of the complex frequency shift in zero field, we are 
interested in its field dependence at fixed temperatures. As an example, we show the field dependence of 
the complex frequency shift for the 10\:nm sample in Fig.~\ref{Fig3}.
From these data one proceeds in two steps. First, the numerical inversion of complex frequency shift 
gives the field dependence of complex effective conductivity. Second, from this $\sigma_{\rm{eff}}$ one 
numerically determines\cite{Mathematica}
parameters $b$ and $\omega_0$ by the use of Eq.~(\ref{eq:4}). The results are given in Fig.~\ref{Fig4}.
One may notice that the depinning frequency is well above the driving frequency for the thinnest sample, 
contrary to the predictions ($\omega_0 \le 100 MHz$) for thick niobium samples.\cite{Gittleman:66} 
We shall return to this point later.

The parameter $b$ is shown in Fig.~\ref{Fig4}a. In the effective conductivity model it is identified
with the volume fraction of the normal cores of the vortices in the mixed state. The field dependence of 
this parameter has nearly linear region at lower fields, as expected for nonoverlapping vortices. 
It has a meaning of reduced field $b=B/B^*_{c2}$ in the model of Bardeen and Stephen,\cite{Bardeen:65} where
$B^*_{c2}$ represents the hypothetical upper critical field given by the equation 
$B^*_{c2}=\Phi_0/(2 \pi \xi(T)^2)$. Therefore, from the linear section of its field dependence one can determine 
the radii of vortices and the GL coherence length at a given temperature.
Detailed analysis of flow resistivity by Larkin and Ovchinnikov leads to the correction factor 0.9 for 
the linear regime, i.e. $b=0.9 B/B^*_{c2}$.\cite{Larkin:86} The extrapolation of this linear section, shown
by the dashed line in Fig.~\ref{Fig4}a, would give $B^*_{c2}(2.9\:\rm K) =1.9 \: \rm T$, 
and $\xi (2.9 \:\rm K) = 41\: \rm{nm}$. 
At higher fields $b$ changes its slope and approaches the normal state value at field $B_{c2}$ lower than 
$B^*_{c2}$. It is the region where vortices start to overlap and the superconducting order parameter is
reduced throughout the sample. The deviation from linearity starts roughly at the field $B=0.63\:B_{c2}$, 
where the distance between vortex centers is 3.4 times the coherence length.

\begin{figure}
\includegraphics[width=8cm]{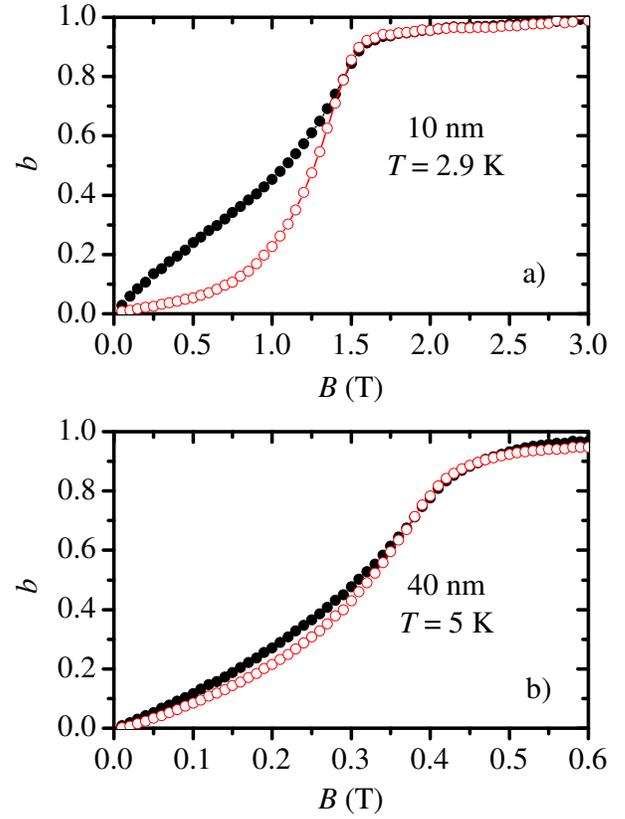}%
\caption{(Color online) Comparison of $b$ obtained by full inversion of complex frequency shift (full black circles) and
$b$ obtained from absorption measurement alone under assumption that depinning frequency is much 
lower than driving frequency (red open circles): a) 10\:nm film at 2.9\:K; b) 40\:nm film at 5\:K.}
\label{Fig5}
\end{figure}
One should mention that numerically determined $b$ values don't always reach unity at $B_{c2}$. 
Close to $B_{c2}$, $b$ looses its meaning as a parameter of the effective conductivity model. 
There are no more well defined vortices and the conductivity is dominated by fluctuations. 
One can still try to numerically obtain $b$ and $\omega_0$, but this leads to negative values of depinning
frequencies, clearly showing the failure of the effective conductivity model in that region.

\begin{figure}
\includegraphics[width=8cm]{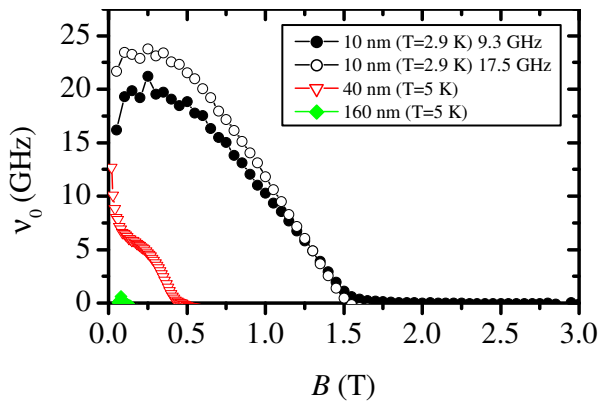}%
\caption{(Color online) Field dependence of the depinning frequency $\nu_0 = \omega_0 / 2 \pi$ obtained by 
driving frequency 9.3\:GHz for three niobium 
films at $T/T_c\approx 0.5$: 10\:nm film at 2.9\:K (black full circles); 40\:nm film at 5\:K (red opened triangles) 
and 160\:nm film at 5\:K (green full diamonds). The depinning frequency obtained by driving frequency 17.5\:GHz 
for the 10\:nm film at 2.9\:K is shown by open black circles.}
\label{Fig6}
\end{figure}

The determination of flow resistivity would not be possible if only the microwave absorption  
had been measured. The effective conductivity depends on two parameters ($b$ and $\omega_0$),
and one has to measure two quantities, absorption and real frequency shift, in order to determine both
parameters. If one supposed that the pinning at microwave frequencies were negligible, one could have determined the
flow resistivity from absorption measurements alone. This would lead to wrong conclusions. 
We show in Fig.~\ref{Fig5}a the comparison of the two ways of reasoning for the 10\:nm film at 2.9\:K. 
The full circles show the parameter $b$ determined from the complex frequency shift
with the depinning frequency as the second result of the numerical inversion. The empty circles are the result of the
numerical inversion of the microwave absorption only, under the assumption that $\omega_0 =0$. One can see that the results 
differ strongly in the low field region, i.e. in the region where the pinning is strong. When the depinning frequency 
falls well below the driving frequency, the two curves overlap.
A similar analysis for the 40\:nm sample is shown in Fig.~\ref{Fig5}b. The two ways of reasoning give closer results since 
the depinning frequency for this sample is lower than the driving frequency 
($\omega_0 \le 0.7 \omega$, see Fig.~\ref{Fig6}).

If one is dealing with the samples whose upper critical field is experimentally unreachable (as in HTSC),
one is lead to infer the upper critical field from the slope of $b$ in low fields. Taking the flow resistivity values 
from microwave absorption, without taking the real frequency shift into account, 
one could reach erroneous values of the upper critical field and GL coherence length. 
This is especially true for samples with high values of depinning frequency, which is usually the case in HTSC.

Bulk metallic superconductors usually don't show pinning at microwave frequencies.\cite{Gittleman:66} 
In Fig.~\ref{Fig6}
we show the depinning frequency of our three niobium films at $T\approx 0.5\:T_c$. The depinning frequency for the 10\:nm 
film at low fields is slightly above 20\:GHz, and for the 40\:nm film it is approximately 7\:GHz. For the 160\:nm film it is below
1\:GHz, i.e. in the frequency range that would be better examined by lower driving frequency. Obviously, in thin films
placed in a perpendicular magnetic field, vortex pinning is dominated by surface pinning centers. Thicker films approach
bulk values of depinning frequencies, showing that the presently grown niobium films don't have bulk pinning centers, 
which are absent also in the best
single crystals.

To check the consistency of depinning frequency values obtained by different driving frequencies, 
we obtained the depinning frequency of 10\:nm sample from measurements in the $_e$TE$_{113}$ mode
with driving frequency of 17.5\:GHz. The result is shown by the open circles in Fig.~\ref{Fig6}, with
satisfactory agreement between the two measurement series.

The above analysis for the 10\:nm film clearly shows linear section of $b$ in moderate fields. Already for the
40\:nm film, the linearity is not perfect. For the 160\:nm film and for single crystal there were no linear sections. 
It is not surprising since the simple Bardeen-Stephen model predicts a linear dependence of $\rho_f/\rho_n$ on $B$
only in the high-$\kappa$ limit. 
Obviously, only the thinnest film considered here fulfills the necessary conditions for the analysis in terms of 
the effective conductivity model.
This analysis is a good model for the analysis of high-$\kappa$ superconductor with high values of upper critical 
fields, typically HTSC.

One can compare the values of $B_{c2}$, obtained by the 90\:\% criterion, with the $B^*_{c2}$ values obtained from the
linear section of field dependence of $b$  by taking into account the Larkin-Ovchinnikov correction $b=0.9\:B/B^*_{c2}$.
\begin{figure}
\includegraphics[width=8cm]{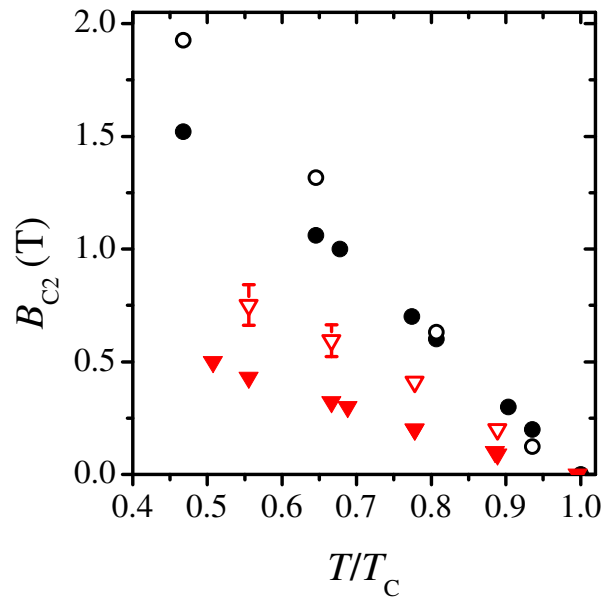}
\caption{(Color online) Upper critical field $B_{c2}$ of the 10\:nm film obtained by the 90\:\% criterion from the microwave absorption
measurements (black full circles). Upper critical field $B^*_{c2}$ of the same sample obtained from the linear section 
of $b$ (black opened circles). The same is shown for the 40\:nm film by red triangles.}
\label{Fig7}
\end{figure}
The comparison is shown in Fig.~\ref{Fig7}. Full circles show $B_{c2}$ obtained by the 90\:\% criterion and opened circles 
show $B^*_{c2}$ values obtained from linear section of $b$. One can see that $B^*_{c2}$ is always slightly higher than 
$B_{c2}$ which would be measured by direct transport or magnetization 
measurement of the crossover to the normal state, if experimentally reachable. The reduction of order parameter due to the 
overlapping of vortices in higher fields is one of the reasons. The other reason for further suppression of the 
actual $B_{c2}$ value can be found in spin paramagnetic effects and spin-orbit scattering.\cite{WHH:66} 
This effect is not significant in our niobium samples since the upper critical fields are low enough, but in HTSC,
one can expect significant suppression of $B_{c2}$ in the fields of the order of 100\:T. 
As a consequence, 
the transition to the normal state is due to a combination of effects, and could not be related uniquely to the coherence length.
However, the microwave determination of flow resistance in low field region gives the best
estimate of the coherence length.

The comparison of $B_{c2}$ and $B^*_{c2}$ for 40\:nm film is also shown in Fig.~\ref{Fig7}. The
$B^*_{c2}$ values are significantly higher  than $B_{c2}$, but one should take this result with caution since there 
is no strictly linear part of $b$ field dependence. This is not surprising since $\kappa\approx 1.5$ doesn't meet 
a high-$\kappa$ condition needed for $B^*_{c2}$ extraction.

\section{Conclusions} \label{sec:level7}

We have measured temperature and magnetic field dependence of the microwave response of high quality niobium films in a
perpendicular static magnetic field. The data have been analyzed by the effective conductivity model for the high-$\kappa$
type-II superconductors in the mixed state.
Our analysis shows that this model yields consistent results when applied in fields not too close to
the transition to the normal state.
Thin niobium films show strong pinning probably due to the surface pinning centers. 
The depinning frequency for the 10\:nm film is $\omega_0 \approx 20\:\rm{GHz}$, and for 
the 40\:nm film it is $\omega_0 \approx 7\:\rm{GHz}$. In thicker samples the depinning frequency is 
lower than $1\:\rm{GHz}$.
One should measure both, the microwave absorption and the fequency shift 
in order to determine flow resistivity
for samples with high values of depinning frequency, as usually found in HTSC.

Although the bulk niobium samples have GL parameter $\kappa$ of the order of unity, very thin niobium films 
have higher $\kappa$-values. They can serve as model system for the upper critical field determination
of high-$\kappa$ superconducting samples. 
In HTSC generally, upper critical fields are of the order of hundreds tesla, and they are unreachable by laboratory magnets. 
Conventional methods for the determination of $B_{c2}$ (transport measurements, susceptibility, torque,...) are based on the 
crossover to the normal state when the applied field equals to $B_{c2}$. The determination of upper critical field is, therefore,
reduced to a narrow temperature region close to $T_c$. The values for lower temperatures are usually determined by an extrapolation.
The correctness of such extrapolations depends on the model used
and there is no assurance that it holds also 
in HTSC. On the other hand, pulsed methods used to reach high applied fields (up to 400 T\cite{Sekitani:04}) are nonequilibrium 
methods and very sensitive to induction, critical currents etc.
In this paper microwave response 
was measured up to the
upper critical field which is for niobium samples reachable by conventional magnets.
The upper critical fields are determined by the flow resistivity in low fields ($B^*_{c2}$) and by the crossover to the normal state ($B_{c2}$).
The two methods give slightly different results. 
The $B_{c2}$ defined as the crossover field to the normal state has certainly great technological importance. However, 
at high fields, typical for HTSC, paramagnetic effects induce depairing and reduce order parameter. If one calculates
the coherence lengths from $B_{c2}$ values obtained in this way, the values are overestimated.
For the considerations of the physical parameters of the superconducting state, it is better to determine the coherence length 
in rather low magnetic fields, deeply in the superconducting state. 
The microwave method tested here is certainly suitable for the determination of $B^*_{c2}$ which gives correct values of effective vortex radii
at a given temperature in the fields low enough that vortices don't overlap. Therefore, the method is relevant for the determination of 
coherence lengths, especially in HTSC.

\acknowledgments 
We thank to Dr. M. Basleti\'c and E. Tafra for the help in the measurements of $\rho_n$.



\begin{thebibliography}{05}
\bibitem{Mathematica}
For numerical determination of $b$ and $\omega_0$ we have used FindRoot routine of commercial software Mathematica (Wolfram Research).
\end{thebibliography}
\end{document}